\documentclass[prl,twocolumn,showpacs,floatfix,amsmath,amssymb,superscriptaddress]{revtex4-1}
\usepackage{amsfonts}
\usepackage{stmaryrd}
\usepackage{bbm}
\usepackage{mathrsfs}
\usepackage{tipa}
\usepackage{amssymb}
\usepackage{txfonts}
\usepackage{graphicx}
\usepackage{dcolumn}
\usepackage{epstopdf}
\usepackage[colorlinks,linkcolor=blue,urlcolor=blue,citecolor=blue]{hyperref}
\usepackage{multirow}
\usepackage{subfigure}
\usepackage{url}

\begin{document}
\newcommand*{\cm}{cm$^{-1}$\,}
\newcommand*{\Tc}{T$_c$\,}

\title{Revealing extremely low energy amplitude modes in the charge-density-wave compound LaAgSb$_2$}
\author{R. Y. Chen}
\affiliation{International Center for Quantum Materials, School of Physics, Peking University, Beijing 100871, China}
\affiliation{Center for Advanced Quantum Studies, Department of Physics, Beijing Normal University, Beijing 100875, China}

\author{S. J. Zhang}
\affiliation{International Center for Quantum Materials, School of
Physics, Peking University, Beijing 100871, China}

\author{M. Y. Zhang}
\affiliation{International Center for Quantum Materials, School of
Physics, Peking University, Beijing 100871, China}

\author{T. Dong}
\affiliation{International Center for Quantum Materials, School of Physics, Peking University, Beijing 100871, China}

\author{N. L. Wang}
\email{nlwang@pku.edu.cn}
\affiliation{International Center for Quantum Materials, School of Physics, Peking University, Beijing 100871, China}
\affiliation{Collaborative Innovation Center of Quantum Matter, Beijing 100871, China}

\begin{abstract}
The layered lanthanum silver antimonide LaAgSb$_2$
was known to experience two charge density (CDW) phase transitions, which were proposed recently to be closely related to the newly identified Dirac cone. We present optical spectroscopy and ultrafast pump probe measurement on the compound. The development of energy gaps were clearly observed below the phase transition temperatures in optical conductivity, which removes most part of the free carrier spectral weight. Time resolved measurement demonstrated the emergence of strong oscillations upon entering the CDW states, which were illuminated to come from the amplitude mode of CDW collective excitations. The frequencies of them are surprisingly low: only 0.12 THz for the CDW order with higher transition temperature and 0.34 THz for the lower one, which shall be caused by their small modulation wave vectors. Furthermore, the amplitude and relaxation time of photoinduced reflectivity stayed unchanged across the two phase transitions, which might be connected to the extremely low energy scales of amplitude modes.
\end{abstract}


\maketitle

Charge-density-waves (CDW) is one of the most fundamental collective quantum phenomena in solids. Charge density waves display periodic modulations of the charge with a period which is commensurate or incommensurate to the underlying lattice.  Most CDW states are driven by the nesting topology of Fermi surfaces (FSs), i.e., the matching of sections of FS to others by a wave vector \textbf{q} = 2\textbf{k}$_F$ , where the electronic susceptibility has a divergence.
A single-particle energy gap opens in the nested regions of the FSs at the transition, which leads to the lowering of the electronic energies of the system. Simultaneously, the phonon mode of acoustic branch becomes softened to zero frequency at \textbf{q} = 2\textbf{k}$_F$ as a result of electron-phonon interaction, which further leads to the periodic modulation of lattice structure.

CDW also has collective excitations referred to as an amplitude mode (AM) and a phase mode. Phase excitation corresponds to the translational motion of the undistorted condensate. In the \textbf{q}=0 limit, the phase mode should locate at zero energy in ideal case since the translational motion does not change the condensation energy\cite{Gruner1988,densitywave}. In reality, due to the presence of impurity or defects, the phase mode is pinned at finite frequency, usually in the microwave frequency range. The pinning/depinning of phase mode has dramatic effect on charge transport properties. By applying dc electric field, the phase mode can be driven into a current-carrying state, leading to nonlinear current-voltage characteristics\cite{MAEDA1985951,PhysRevLett.74.5264,PhysRevLett.42.1423}. On the other hand, the amplitude mode involves the ionic displacement and has a finite energy even at \textbf{q}=0 limit. For most CDW materials, the amplitude mode has an energy scale of about 10 meV (or $\sim$ 2 THz)\cite{Tsang1976,PhysRevLett.83.800,PhysRevB.66.041101,PhysRevLett.101.246402,PhysRevB.78.201101,Porer2013,1347-4065-46-2R-870}. Due to presence of such a gap for the amplitude mode (i.e. the mode energy at \textbf{q}=0), its effect on low temperature physical properties of CDW condensate has been much less studied.
Generally, amplitude modes can be treated as optical phonons, reflecting the oscillations of ions. Nevertheless, it is also proposed to be intimately related with the modulation of CDW energy gap, which is the most fundamental difference between AM and optical phonons.

\begin{figure}[htbp]
  \centering
  \includegraphics[width=7.5cm]{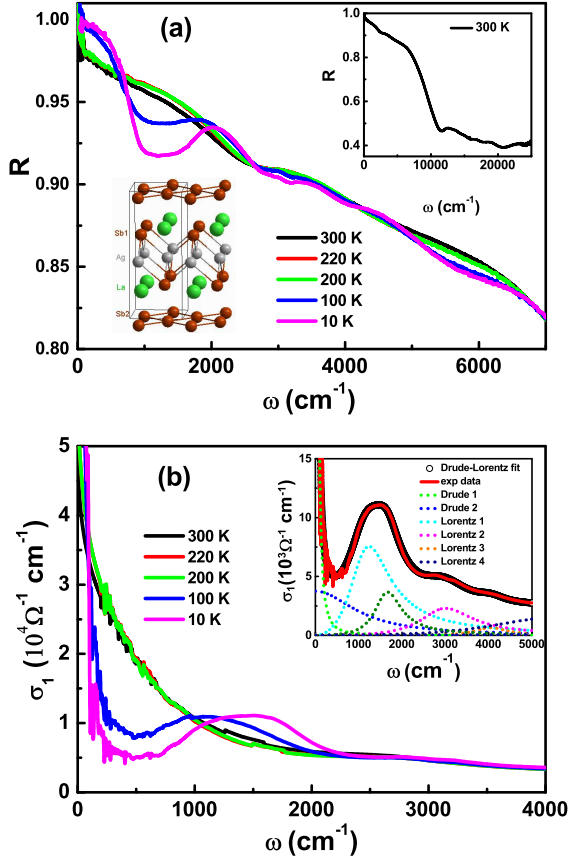}\\
  \caption{The temperature dependent (a) reflectivity $R(\omega)$ and (b) optical conductivity $\sigma_1(\omega)$ for LaAgSb$_2$ single crystal below 7000 \cm and 4000 \cm, respectively. The insets of (a) present crystal structure and $R(\omega)$ at 300 K over a broad frequency range up to 25 000 \cm. The inset of (b) displays $\sigma_1(\omega)$ at 10 K together with a Drude-Lorentz fit.}\label{Fig:ref-cond}
\end{figure}

The compound LaAgSb$_2$, which crystallizes in simple tetragonal ZrCuSi$_2$ structures as shown in the lower inset of Fig. \ref{Fig:ref-cond} (a),
was revealed to experience two CDW phase transitions around $T_{C1}$ = 207 K and $T_{C2}$ = 186 K, providing a unique platform to study the effect of amplitude modes on physical properties, as will be elaborated below.
The modulation wave vectors corresponding to the CDW orderings were identified to be $\sim$ (0.026 0 0) for the higher temperature transition and (0 0 0.16) for the lower one, by X-ray diffraction\cite{Song2003}.
The electronic band structure calculation reveals that there are four bands crossing the Fermi energy $E_F$, being consistent with de Haas-van Alphen (dHvA) measurement\cite{Myers1999b}, and the two CDW modulations are dominantly driven by the FS nesting\cite{Song2003}.

Recently, some renewed interest have been triggered in the LaAgSb$_2$ compound because of the possibility of hosting Dirac fermions in this material. First-principles electronic structure calculation demonstrated that there are both linear and parabolic bands crossing the Fermi level, which was supported by the appearance of large linear magnetoresistance and further magnetothermopower measurements\cite{Wang2012d}. In addition, angle-resolved photoemission spectroscopy (ARPES) directly observed the Dirac-cone like structure in the vicinity of the Fermi level, which was formed by the crossing of two linear Sb 5$p_{x,y}$ energy bands\cite{Shi2016}. Of most significance, the nested FS associated with CDW phase transition was proved to emerge from two almost paralleled segments of the Dirac cone, leading to a very small modulation wave vector 2\textbf{k}$_F$.

To investigate the potential relationship between Dirac fermions and CDW orderings, and to gain insight into the possible effect of AM, we performed infrared spectroscopy and ultrafast pump probe measurement on the single crystalline LaAgSb$_2$, both of which are supposed to be very sensitive to the formation of symmetry breaking gaps. Particularly, the ultrafast spectroscopy are playing a more important role than ever in studying the coherent vibrational dynamics, since this method can easily access to oscillations of extremely low energy scale, as a stretching of neutron and Raman scattering.

We have successfully synthesized large pieces of LaAgSb$_2$ single crystals by self flux method, as described in previous report\cite{Myers1999}. The plate like samples have shiny surfaces after eliminating the extra flux.
The in plane reflectivity $R(\omega)$ was measured by the combination of Fourier transform infrared spectrometer Bruker 113 V and 80 V in the frequency range from 30 to 25 000 \cm.

As displayed in the upper inset of Fig.\ref{Fig:ref-cond}(a), the room temperature reflectivity was characterized by typical metallic responses: $R(\omega)$ approaches unit at zero frequency and a well defined plasma edge was clearly observed around 10 000 \cm. The high plasma edge frequency indicates that the density of free carriers in this material is rather large.
The main panel of Fig.\ref{Fig:ref-cond}(a) shows the reflectivity below 7000 \cm at several selected temperatures. Above
$T_{C1}$, $R(\omega)$ increases monotonically as frequency deceases and the low energy part increases slightly upon cooling, both of which belong to simple metallic behaviors. The reflectivity spectrum of 200 K almost overlaps with that of 220 K in the normal state. Although the first CDW instability has already set in at 200 K, the modification of the electronic band structure is yet too weak to be detected. With temperature further cooling far below $T_{C2}$, a pronounced dip structure appears roughly between 1000-2000 \cm, which strongly evidences the formation of a charge gap in the vicinity of Fermi level, due to the development of the CDW order. This dip feature grows more dramatic as temperature decreases and its center frequency keeps shifting to higher energy at the same time, indicating the continuous enhancement of the CDW gap. In the meantime, the low energy reflectivity gets even higher than the normal metallic state.

The real part of optical conductivity $\sigma_1(\omega)$ was derived from $R(\omega)$ through Kramers-Kronig transformation, as shown in Fig.\ref{Fig:ref-cond} (b). The low energy part of $R(\omega)$ was extrapolated to unit by Hagen-Rubens relation, and for the high frequency extrapolation we have employed the x-ray atomic scattering functions method\cite{PhysRevB.91.035123}. In the normal state, the optical conductivity exhibits clear Drude peaks centered at zero frequency, and the broad half width indicates large scattering rate $\gamma$ of the free carriers. Upon entering the CDW state, the spectral weight of the Drude peak was substantially removed and transferred to higher energies, and a broad bump locating around 1000 \cm emerges as a consequence.
Moreover, the bump position shifted to higher energies as temperature decreases. In fact, these features are the spectroscopic fingerprints of density wave materials, since the case-I coherence factor of density wave order is expected to cause a sharp rise in the optical conductivity spectrum just above the energy gap. Here, we can roughly take the central position of the emerging bump as the energy scale of the CDW gap.

\begin{figure*}[hbtp]
  \centering
  \includegraphics[width=5.3cm]{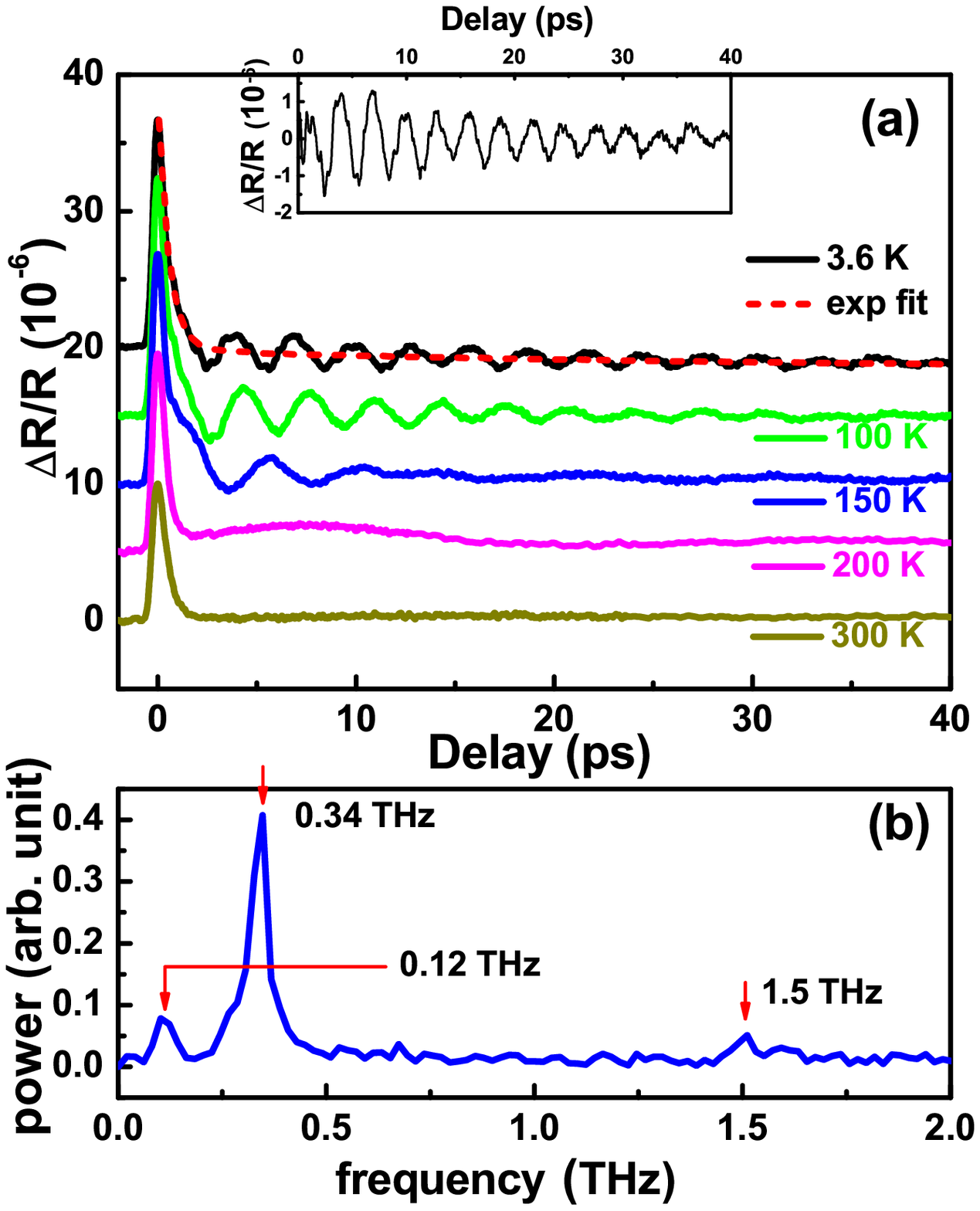}
  \includegraphics[width=5.8cm]{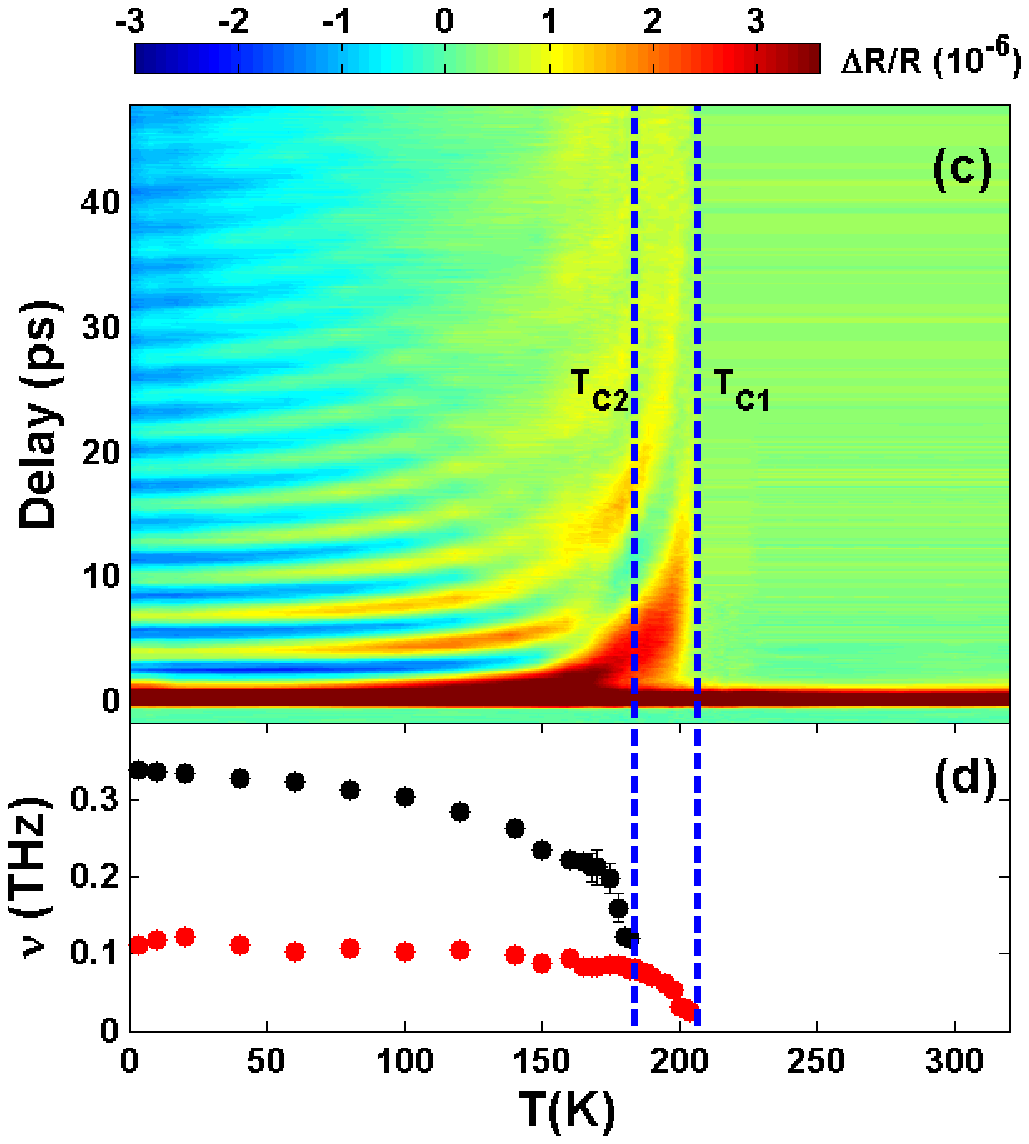}
  \includegraphics[width=4.6cm]{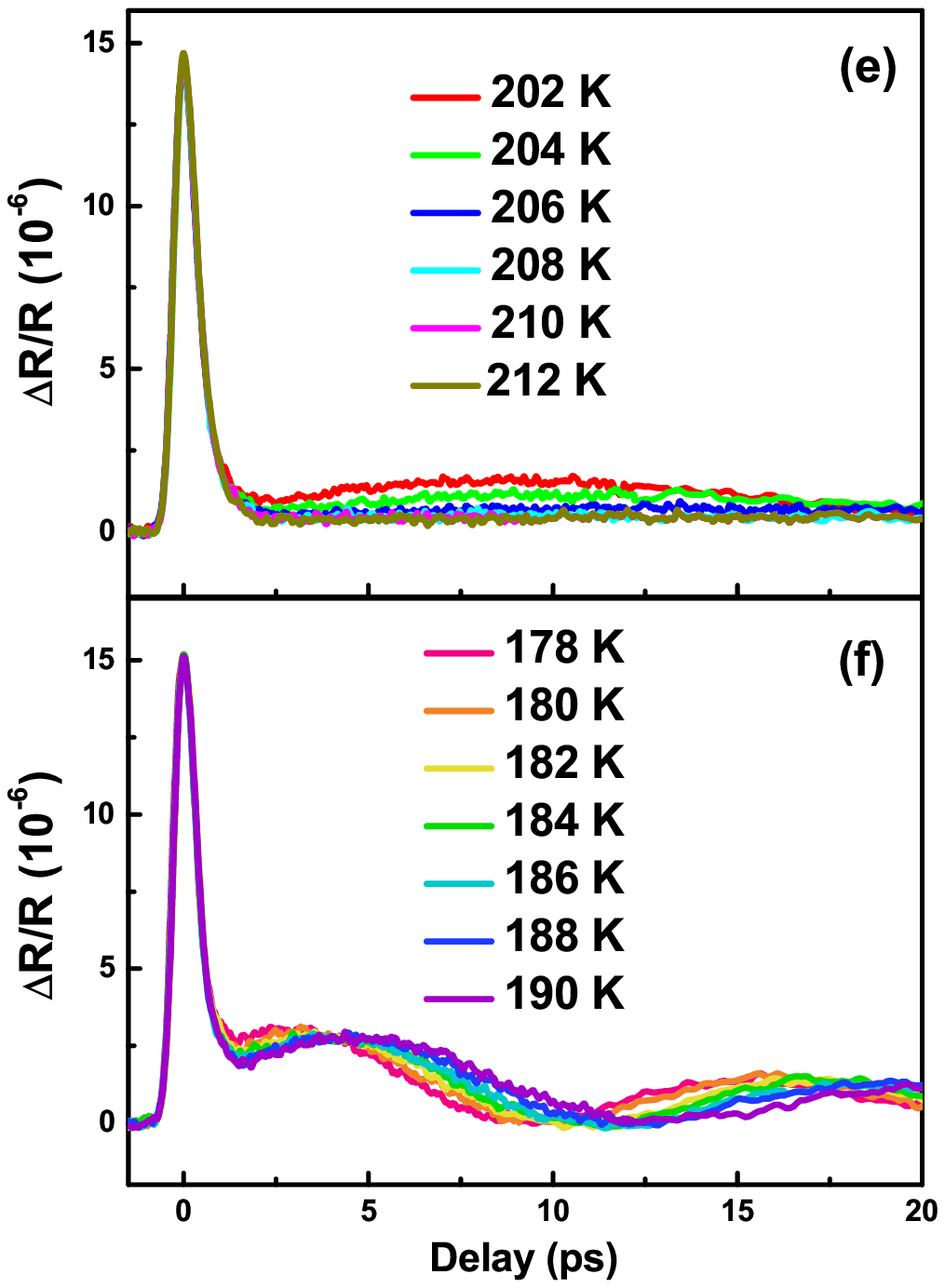}\\
  \caption{(a) The photoinduced reflectivity $\Delta R/R$ as a function of time delay at several selected temperatures. The experimental data are shifted upwards for the sake of clearness. The red dotted line is the exponential fitting of $\Delta R/R$ at 3.6 K. The inset shows the oscillation part of $\Delta R/R$ as described in the main text. (b) The fast Fourier transformation of the oscillation part of $\Delta R/R$ at 3.6 K. (c) The photoinduced reflectivity $\Delta R/R$ as a function of time delay and temperature. (d) The temperature dependent frequencies of the two CDW amplitude modes $\nu_1$ (red dot) and $\nu_2$ (black dot). (e) The photoinduced reflectivity $\Delta R/R$ across the first CDW phase transition and (f) the second CDW phase transition. }\label{Fig:DeltaR}
\end{figure*}

To make the estimation more accurate, we have employed the Drude-Lorentz model to fit the optical conductivity.
\begin{equation*}
\epsilon(\omega)= \epsilon_{\infty}-\sum_{s}{\frac{\omega_{ps}^2}{\omega^2+i\omega/\tau_{Ds}}}+ \sum_{j}{\frac{S_j^2 }{\omega_j^2-\omega^2-i\omega/\tau_j}}. 
\end{equation*}
Here, $\varepsilon_{\infty}$ stands for the dielectric constant at high energy; the middle and last terms are the Drude and Lorentz components, respectively. We found that the $\sigma_1(\omega)$ can be well reproduced by two Drude terms and several Lorentz terms.
The Drude term is used to describe the response of itinerate carriers, whereas the Lorentz term stands for interband transitions and excitations across energy gaps. It is found that the bump appearing at low temperature is too broad to be reproduced by a single Lorentz term, which demonstrates that there are at least two different gaps opening associating with the CDW phase transitions. According to our experimental configuration, the obtained $\sigma_1(\omega)$ is supposed to be paralleled with $ab$ plane, thus the two gaps shall be both originated from the CDW order with higher transition temperature. It is reasonable because the specific wave vector connects different paralleled regions of the FS. The obtained gap energies are about 1240 \cm ($\sim$ 154 meV) and 1670 \cm ($\sim$ 207 meV) respectively. It is worth noting that the Drude peaks remain in the CDW state, which agrees well with the high reflectivity near zero frequency, indicating that the FS are partially gapped by the CDW orders. The number of lost free carriers could be estimated by the variation of plasma frequency $\omega_p\sim\sqrt{n/m^*}$, where $n$ and $m^*$ represent the number and effective mass of free carriers respectively.
As we have employed two Drude terms to fit $\sigma_1(\omega)$, the overall plasma frequency was obtained by $\omega_p=(\omega_{p1}^2+\omega_{p2}^2)^{1/2}$, where $\omega_{p1}$ and $\omega_{p2}$ represent the individual plasma frequencies corresponding to respective Drude component. According to our fitting procedure, as presented in the inset of Fig. \ref{Fig:ref-cond} (b), $\omega_p$ varies from 40400 \cm at room temperature to 25500 \cm at 10 K, which indicates that about 60 \% of free carriers are removed by the opening of CDW gaps.
Meanwhile, the scattering rate $\gamma_1$ decrease violently from 457 \cm to 14 \cm, and $\gamma_2$ also reduces from 2073 \cm to 1097\cm, evidenced by the narrowing of Drude peaks, which makes the compound an even better metal in spite of the substantial carrier density loss.

Some more interesting properties of this material have been revealed by our ultrafast pump probe measurement, which has been proved to be particularly powerful in detecting both of the single particle excitations across small energy gaps\cite{PhysRevLett.82.4918,PhysRevLett.104.027003,PhysRevB.84.174412,Chen2014a} and collective excitations relevant to quantum orderings\cite{PhysRevLett.101.246402,Albrecht1992,PhysRevLett.111.057402}. We used a Ti:sapphire oscillator as the light source for both pump and probe beams, which can produce 800 nm pulsed laser at 80 MHz repetition. The 100 femtosecond time duration of the laser pulses enables ultrashort time resolved measurement. The fluence of the pump beam is about 2 $\mu J/cm^2$, and the fluence of the probe beam is ten times lower. In order to eliminate the noise caused by stray light, the pump and probe pulses were set to be cross polarized and an extra polarizer was mounted just before the detector.

The photoinduced transient reflectivity $\Delta R/R$ at several selected temperatures are displayed in the main panel of Fig.\ref{Fig:DeltaR} (a), as a function of time delay. At room temperatures, $\Delta R/R$ initially increases in a very short time due to the pump excitation, then drops rapidly back to the equilibrium state within picoseconds, the falling slope of which could be well described by a single exponential decay $\Delta R/R= A ex p (-t/\tau)$, where $A$ represent the amplitude of the photoinduced reflectivity and $\tau$ stand for the relaxation time of the decay. Remarkably, pronounced oscillatory signals were observed at lower temperatures, the periodic time of which changes dramatically with temperature variation. In order to analyze the oscillatory component quantitatively, we first subtracted off the exponential fitting part, as shown in Fig.\ref{Fig:DeltaR} (a) by red dotted line for $T=3.6$ K. The fast Fourier transformation of the residual part, as shown in the inset of this figure, demonstrated the frequencies of the oscillations. It is clearly seen in Fig.\ref{Fig:DeltaR}
(b) that there are three distinct modes at 3.6 K, the frequencies of which are about 0.12 THz, 0.34 THz, and 1.5 THz, respectively. It should be noted that the 0.12 THz oscillation is out of the resolution limit of neutron and Raman scattering.

In order to investigate the origination of these oscillations, we have explored their temperature dependence thoroughly from 3.6 K to 320 K. The raw experimental data for all the measured temperature range were plotted in a psudo-color picture, as shown in Fig.\ref{Fig:DeltaR} (c). The oscillatory components, which show up abruptly just below $T_{C1}$, were highlighted by setting the corlorbar in a small amplitude scale. Although we can not resolve all the three components, the periodic time of the observed oscillation in real time grows longer continuously by approaching the first CDW phase transition temperature from below, suggesting that this oscillation is very likely to be related to collective modes of the long range CDW order.


The frequencies evolution of the oscillations as a function of temperature were displayed in Fig.\ref{Fig:DeltaR}(d), yielded by fast Fourier transformation. The one locates around 1.5 THz only exists at very low temperatures, so we did not show it in this figure. The modes $\nu_1=$ 0.12 THz is almost temperature independent at low temperature, but exhibits dramatic softening on approaching $T_{C1}$. It proceeds in a fashion very similar with the BCS gap function, indicating that $\nu_1$ is probably the amplitude mode of the first CDW ordering. Analogously, $\nu_2=$ 0.34 THz softens significantly around $T_{C2}$, implying the appearance of another amplitude mode. Nevertheless, such oscillation observed in time domain could also be connected to coherence lattice vibrations, such as non-equilibrium phonon dynamics, which may emerges upon entering the CDW state due to lattice distortion accompanied with the phase transition\cite{Schafer2010}. Furthermore, the surprisingly low energy scales of both $\nu_1$ and $\nu_2$ are rarely seen for amplitude mode of CDW orderings, whose magnitude is usually one or a few terahertz. On the contrary, the frequencies of acoustic phonon and phase mode of the CDW ordering locate generally in the sub-THz area\cite{Lim2005,Ren2004}, being consistent with our experimental results.

\begin{figure}[htbp]
  \centering
  \includegraphics[width=8cm]{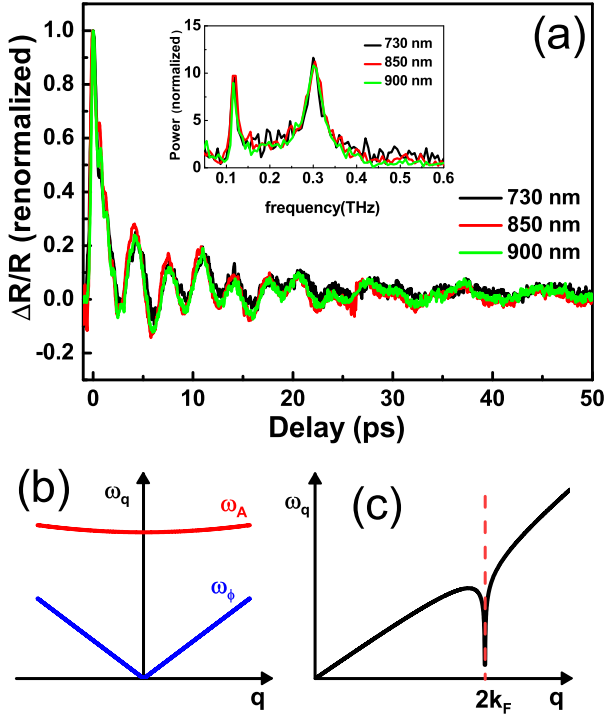}\\
  \caption{(a) The renormalized photoinduced reflectivity obtained by different pump and probe wavelength at 100 K. The inset displays the fast Fourier transformation of the oscillation part corresponding to different wavelength. (b) Amplitude and phase mode dispersion relations. (c) Softening of acoustic phonon mode at CDW wave vector 2\textbf{k}$_F$. }\label{Fig:wave}
\end{figure}


To exclude the possibility of coherent acoustic phonon and phason, we further performed wavelength dependent measurement at a fixed temperature of 100 K. The wavelength of both pump and probe pulses were tuned to 730 nm, 850 nm and 900 nm separately, and the obtained photoinduced signals are shown in the main panel of Fig.\ref{Fig:wave} (a). There are actually no obvious differences between those signals corresponding to different wavelengthes. In addition, the individual frequencies obtained through a standard procedure described above was displayed in the inset of Fig.\ref{Fig:wave} (a), all of which indicates two modes locating around 0.12 THz and 0.3 THz, respectively. It is well known that the frequency of amplitude mode exhibit a gap when the wave vector approaches zero, whereas the phase mode is gapless, as shown in Fig.\ref{Fig:wave} (b). The dispersion of acoustic phonon is similar to that of phase mode. That is, the frequency of acoustic phonon shall increase with the probe wavelength decreasing\cite{Ren2004}.
Consequently, the contribution from acousticlike phonons including phason for the modes observed in our experiments could be ruled out immediately. Although optical phonons of similar energy scale have been identified in other time-resolved measurement\cite{He2016}, the temperature dependent evolution behaves quite differently.

Now that we are quite certain $\nu_1$ and $\nu_2$ come from the amplitude mode of the two CDW orderings, another question springing to mind is that the frequencies of $\nu_1$ and $\nu_2$ seem to be extraordinarily low. It is well known that the acoustic phonon softens significantly at \textbf{q} = 2\textbf{k}$_F$  upon the CDW phase transition, as exhibited in Fig.\ref{Fig:wave} (c). Theoretical calculation suggested that the frequency of amplitude mode $\omega_A$ satisfies\cite{densitywave}:
\begin{equation}\label{eq:omegaA}
 \omega_A=(\lambda\omega_{2k_F}^2+\frac{1}{3}\frac{m}{m^*}v_F^2q^2)^{1/2},
\end{equation}
where $\omega_{2k_F}$ is the frequency of the softening phonon above the phase transition temperature; $\lambda$ and $V_F$ are the electron-phonon coupling constant and Fermi velocity, respectively. In the $q=0$ limit, $\omega_A\sim \lambda^{1/2}\omega_{2k_F}$. The modulation wave vectors of the two CDW order in LaAgSb$_2$ were identified to be $\sim$ (0.026 0 0) for the higher temperature transition and (0 0 0.16) for the lower one\cite{Song2003}, both of which are uncommonly small, suggesting extremely long lattice modulation periods in the real space. As the softened phonon dispersion is of acoustic type,
the tiny modulation wave vectors directly lead to a low energy $\omega_{2k_F}$, which further give rise to rarely low energy amplitude modes $\nu_1$ and $\nu_2$. It is worth pointing out that  the frequency of $\nu_1$ is lower than $\nu_2$, which is in good agreement with our interpretation, because the modulation wave vector of the first phase transition is much smaller than the second one.

As mentioned above, the ultrafast pump probe measurement is very powerful in detecting the low energy gaps opening on the FS, such as density wave or superconducting gaps, which can be well explained by the phenomenological Rothwarth-Taylor (RT) model\cite{PhysRevLett.19.27}. The two most pronounced features of relevant phase transitions are the significant enhancement of the amplitude $A$ and the quasi-divergence behavior of the relaxation time $\tau$\cite{PhysRevLett.82.4918,PhysRevLett.104.027003,PhysRevB.84.174412,Chen2014a}. However, we did not observe either of such characters in LaAgSb$_2$ single crystals. As plotted in Fig. \ref{Fig:DeltaR} (e) and (f), the single exponential part of $\Delta R/R$ remains unchanged across both of the two phase transitions, generating constant amplitude $A$ and relaxation time $\tau$. It seems that the CDW phase transitions do not have any effect on the transient reflectivity, and thus physical properties of the compound. To our knowledge, such an exotic phenomenon has never been observed ever before, and it is hard to understand considering that more than half of the FS are gapped by the CDW orders. Compared with other CDW materials, we highly suspect that the extraordinarily low energy AM may play some role in this case. Based on previous time-resolved ARPES measurement, the AM of CDW orders could induce periodic modulation of the conduction band, which further lead to CDW gap modulation\cite{Schmitt2008,Liu2013}. Around the phase transition temperatures where the energy scales of gaps are actually very small, such modulation may probably melt the CDW gap within the time scale of the AM period, which is exceptional long for LaAgSb$_2$. Consequently, the photoinduced $\Delta R/R$ does not change across the CDW phase transitions since the relaxation time is only about 0.5 ps. Although there might be other possible mechanisms, we believe that the AM, which couples to lattice and behaves like lattice vibration (i.e. phonon), are most relevant.

In summary, we have utilized infrared and ultrafast pump probe spectroscopy to investigate the charge and coherent dynamics of the LaAgSb$_2$ single crystal. The optical conductivity revealed two energy gaps related to the higher temperature CDW phase transition, whose energy scales were identified to be 154 meV and 207 meV respectively. More significantly, our time resolved ultrafast measurement revealed two collective amplitude mode oscillations at surprisingly low energy scales of about 0.12 THz and 0.34 THz at the lowest temperature.
This low energy scale implies that acoustic phonon mode, which softens and triggers CDW instability, also has very low energy scale. We elaborate that those unusual properties are closely linked to the extremely small nesting wave vectors of the two CDW orders, which possibly result in the bizarre transient dynamics of the compound.
\begin{center}
\small{\textbf{ACKNOWLEDGMENTS}}
\end{center}

We acknowledge useful help from Dr. H. P. Wang in the data analysis. This work was supported by the National
Science Foundation of China (No. 11327806), and the National Key Research and Development Program of China (No.2016YFA0300902, and No.2016YFA0302300).

\bibliographystyle{apsrev4-1}
  \bibliography{LaAgSb2}

\end{document}